# A Voice Disease Detection System Based on MFCCs and Single-Layer CNN

Xiaoping Xie, Hao Cai, Can Li,

***Abstract*—The incidence rate of voice diseases is increasing year by year. The use of software for remote diagnosis is a technical development trend and has important practical value. Among voice diseases, common diseases that cause hoarseness include spasmodic dysphonia, vocal cord paralysis, vocal nodule, and vocal cord polyp. This paper presents a voice disease detection method that can be applied in a wide range of clinical. We cooperated with Xiangya Hospital of Central South University to collect voice samples from sixty-one different patients. The MFCC parameters are extracted as input features to describe the voice in the form of data. An innovative model combining MFCC parameters and single convolution layer CNN is proposed for fast calculation and classification. The highest accuracy we achieved was 92%, it is fully ahead of the original research results and international advanced. And we use Advanced Voice Function Assessment Databases (AVFAD) to evaluate the generalization ability of the system we proposed, which achieved an accuracy rate of 98%. Experiments on clinical and standard datasets show that for the pathological detection of voice diseases, our method has greatly improved in accuracy and computational efficiency.**

## I. INTRODUCTION

ABOUT 3%-9% of people in the world have voice diseases or voice disease symptoms of different degrees. The voice disease affects the normal communication, life, and work quality of human beings [1]. Several factors lead to voice pathology such as extreme damage of certain organs, air pollution, smoking, eating habits and stress [2]. Especially for those who need to speak loudly and frequently for long-term professional reasons, such as teachers, singers, auctioneers, lawyers and actors, the incidence of voice disease is higher [3]. Clinically, the location of laryngeal lesions is always determined by laryngeal endoscopy and diagnose laryngeal muscle and vocal cord diseases by laryngeal electromyography. These two clinical methods of voice disease detection depend to a considerable extent on the clinical experience and subjective judgment of doctors, also it incurs higher cost [4]. There is an intuitive biological corresponding relationship between the speech conditions and acoustic characteristics of healthy and various kinds of morbid sounds, which also provides a physical explanation for the speech waveforms generated and transmitted by sound organs. The quality of voice can be objectively and scientifically evaluated through the extraction and processing of voice signal characteristics of patients [5]. As a computer-aided pathological assessment, we can start with several aspects, including perceptual assessment, acoustic assessment, aerodynamic assessment, endoscopic imaging, which can be combined with machine learning methods to predict and evaluate from the point of view of data [6]. At the same time, these methods also avoid secondary injury to the patient caused by laryngoscopy and help doctors diagnose the voice diseases of the patients at a relatively cheap cost [7].

The generation of voice can be described by the following process. First, the lungs breathe, and the airflow causes the vocal cords to vibrate regularly to produce sound waves, and finally passes through the mouth and nose. Usually, patients with voice diseases sound stressful, stiff, or extremely weak and wheezing. In other words, it is accompanied by poor sound quality. Among the common voice diseases, the diseases that lead to dysphonia include spasmodic dysphonia, vocal cord paralysis, vocal cord nodules and vocal cord polyps.

Spasmodic dysphonia is the neurological voice disorder that arises due to tightening of the vocal folds, in people with spasmodic dysphonia there exists involuntary tightening of the vocal folds by affecting the voice [8]. Vocal cord paralysis is mainly caused by lesions, invasion, compression and other injuries of the surrounding tissues of the vagus nerve and the recurrent laryngeal nerve. These result in vocal cord movement disorder. The clinical manifestations are hoarseness, weakness and even aphasia [9]. Vocal cord nodules are a special type of

---

This paragraph of the first footnote will contain the date on which you submitted your paper for review, which is populated by IEEE. It is IEEE style to display support information, including sponsor and financial support acknowledgment, here and not in an acknowledgment section at the end of the article. For example, "This work was supported in part by the U.S. Department of Commerce under Grant 123456." The name of the corresponding author appears after the financial information, e.g. *(Corresponding author: M. Smith).* Here you may also indicate if authors contributed equally or if there are co-first authors.

The next few paragraphs should contain the authors' current affiliations, including current address and e-mail. For example, First A. Author is with the National Institute of Standards and Technology, Boulder, CO 80305 USA (e-mail: author@ boulder.nist.gov).

Second B. Author was with Rice University, Houston, TX 77005 USA. He is now with the Department of Physics, Colorado State University, Fort Collins, CO 80523 USA (e-mail: author@lamar.colostate.edu).

Third C. Author Jr. is with the Electrical Engineering Department, University of Colorado, Boulder, CO 80309 USA, on leave from the National Research Institute for Metals, Tsukuba 305-0047, Japan (e-mail: author@nrim.go.jp).

Mentions of supplemental materials and animal/human rights statements can be included here.

Color versions of one or more of the figures in this article are available online at http://ieeexplore.ieee.org

chronic laryngitis, which can be defined as small, benign thickenings of the margins of the vocal cords. The main clinical symptom is rough voice, often hoarseness at high pitch, accompanied by delayed pronunciation and timbre change [10]. Vocal cord polyp is a benign proliferative lesion occurring in the superficial layer of the vocal cord proper, and the main clinical symptom is hoarseness.

Researchers often conduct algorithm research on predictive classification of one or two of the diseases. The accuracy of unitary and binary classification has generally reached more than 90%, but for four-category research, the accuracy is less than 80% [11][12].

In the research introduced in this paper, we proposed a method based on deep learning to effectively distinguish four diseases mentioned above. As the paper unfolds, Section II presents related work. Section III gives a description of the dataset. Section IV describes the methodology which consists of data pretreatment, feature extraction and algorithm model. Section V shows our experimental tests and results, in this section, we use several evaluation indicators to test the predictive classification effect of the system under clinical audio. Finally, the effect of the model is discussed by using a more convincing dataset in Section VI.

## II. RELATED WORK

### A. Datasets

Massachusetts Eye and Ear Infirmary Database (MEEI) is one of the most used databases which consists of various recordings of organic, neuralgic, traumatic, and psychogenic voice disorders. But MEEI Database records voice samples from different environments and frequencies, which may be defects for researchers when study intensively [13]. AVPD (Arabic Voice Pathology Database) collected samples by recording three vowels, running speech and isolated words of subjects. Especially, it provides perceptual severity of each sample [14]. SVD (Saarbrucken Voice Database) contains subjects' voice samples who suffer from several pathologies, and the pathologies' recordings can be classified into functional and organic [15]. VOICED is a database focused on automatic voice disorder detection and classification but has a drawback that it consists of smaller samples than other databases. AVFAD is an open access resources based on a sample of 346 patients clinically diagnosed with vocal pathology and 363 individuals with no vocal alterations. It contains sustaining vowels /a, i, u/, reading of sentences, reading of balanced text and spontaneous speech [16].

### B. Features

Features need to be extracted for pattern recognition. A group of acoustic characteristic parameters which can effectively express the characteristics of pathological voice can make the prediction results more accurate. L. Salhi and others realized the classification of normal and pathological speech by using traditional acoustic features such as pitch, formant, and wavelet coefficients [17]. A. Leonardo, M. Forero and others extract parameters of time and frequency domain, and parameters represent the variations of the fundamental frequency from glottic signal [18]. Ghulam Muhammad and others applied the audio features of MPEG-7 to the speech classification of asthma diseases and achieved satisfactory results [19]. And some extract prosodic vocal tract and excitation features [12]. To enhance the richness of features, some scholars have constructed and trimmed a salient feature set containing different features, such as perturbation features, spectral and cepstral features, complexity features [20]. In [21-25], MFCC (Mel-Frequency Cepstral Coefficients) are widely used in voice signal feature extraction and have achieved superior results.

### C. Algorithms

Decades of research has proved that the application of machine learning method (SVM, GMM, ELM) in the classification of health and disease speech is effective and accurate. Researchers studied and applied various Deep learning architectures which include a 5-layer network, 5-layer CNN and RNN. In the experiment, 5-layer CNN got a satisfying result, which received 96% sensitivity and 18% specificity [26]. The sensitivity and specificity mentioned here are related to the recall and precision in the medical field. Sensitivity, figuratively speaking, is given a series of samples, how many proportions of patients can be judged; specificity is the ability to determine a certain disease. In [12] a Deep belief network has been proposed to extract feature with SVM to accomplish the classification. Also, some researchers built and improved a DNN model, which includes 5 hidden layers with 200 neurons. Balanced the sensitivity and specificity by adjusting parameters and achieved a result of 93.1% and 46% [21]. Traditional machine learning models (such as Support Vector Machine, Random Forest, K-Nearest Neighbor, and Gradient Boosting) have also achieved reliable results in classification [23][25]. For four-category problems, researchers proposed convolutional architectures combined with feed-forward neural networks, to classify four types of voice disorders and achieved 57% accuracy [27].

## III. MATERIALS

### A. Clinical samples

After seeking the consent of patients and their families, we selected the relevant patients from Xiangya Hospital from January 2021 to December 2022 as the study subjects. The patients were diagnosed by doctors with modern medical means such as medical history inquiry, physical examination, and electronic laryngoscopy. We look forward to designing and testing our voice disease surveillance system first with the goal of distinguishing the specific voice audio in these patients.

The patients, accompanied and guided by the medical staff, vocalize the designated phonemes. Since the case population is all adult Chinese, we propose that the phonemes of vocalization are Chinese vowels /a, o, e, i, u, ü/ and /wo, ei/.

In most databases or study, researchers focus on continuous pronunciation of one vowel(/a/) or three combinations (/a, i, u/) [29][30]. Objectively speaking, these only examine one aspect of speech, and we increase it to 8 phonemes, which can avoid the limitations of most speech pathology detection strategies. During voice capture, we used relatively simple acquisition equipment such as recording pens, mobile phones, etc. for recording. We hoped that the audio quality of our experiments is as bad as possible, to fit the actual clinical environment and equipment conditions.

Finally, the patient's audio dataset we created includes 1464 voice audio strips of patients with four diseases, each of which is about 3 seconds. The dataset contains files as follow:

1) Audio files
   Each of the eight phonemes was read three times for about 3 seconds, the detailed distribution of audio files is shown in Table I.
2) Case Reports
   Hospital diagnosis records include doctors' clinical judgments and pathological analysis, patients' physiological information and daily life habits, etc.

Table I.
Distribution of audio files

| Classes | /a/ | /o/ | /e/ | /i/ | /u/ | /ü/ | /w/ | /ei/ |
|---|---|---|---|---|---|---|---|---|
| Spasmodic dysphonia | 36 | 36 | 36 | 36 | 36 | 36 | 36 | 36 |
| Vocal cord paralysis | 54 | 54 | 54 | 54 | 54 | 54 | 54 | 54 |
| Vocal cord nodules | 33 | 33 | 33 | 33 | 33 | 33 | 33 | 33 |
| Vocal cord polyps | 60 | 60 | 60 | 60 | 60 | 60 | 60 | 60 |

## IV. METHOD DESIGN BASED ON MFCC AND CNN

The initial stage of almost all automatic speech recognition systems is to identify audio components that are useful for content detection or feature extraction. The research work in this part of this paper consists of three parts as Fig. 1 shows. First, the collected audio signal is preprocessed in wav format; then the features of the preprocessed audio signal are extracted; finally, an algorithm model is established for training and testing.

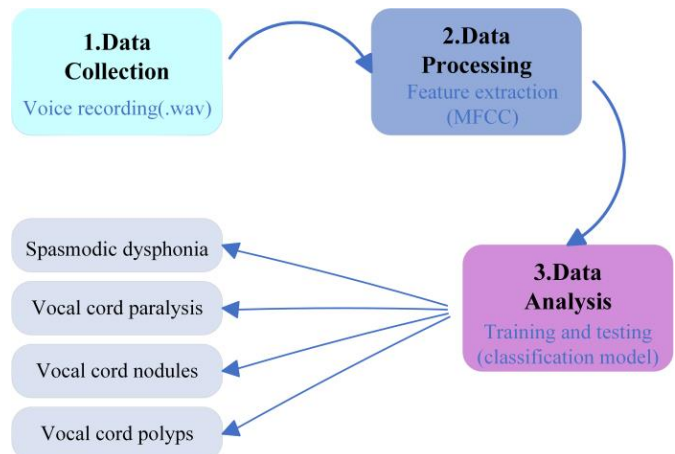

Fig. 1. The block diagram for the pathological voice classification

### A. Audio Pre-processing

The collected vowel audio, whose energy distribution is mainly concentrated below 1kHz. And the signal in high frequency band has negligible impact on the study, so we skipped the step of pre-emphasis. The preprocessing process in this paper consists of two steps, namely Voice Activity Detection and Audio Clipping.

Considering the objective factors of the acquisition process, the patient's recording is usually an extended period of audio, which contains some useless information. Voice activity detection can find the starting and ending point of the effective sound segment in the signal data, which facilitates the interception of audio clips. Finally, we intercept the effective audio into several short audio data with a length of 0.5s, each copy contains only one phoneme.

### B. Feature Extraction

In the feature extraction stage, we referred to the practice in [27][31]. MFCCs and Mel-spectrogram have a favorable effect in characterizing speech signal information. MFCC simulates, to a certain extent, the perceptual characteristics of the human ear to sound. The auditory characteristics of the human ear are consistent with the increase of Mel frequency. It shows a linear distribution with the actual frequency below 1000Hz, and a logarithmic increase above 1000Hz.

The extraction process of MFCC is shown in Fig. 2. The entire process can be described as follows. After the voice signal in wav format is input, pre-emphasis, framing, windowing, and FFT are performed on it. This operation discretizes the continuous voice waveform and converts it into a one-dimensional array of large values. Then a set of Mel-scale triangular bandpass filters is passed. Then perform logarithmic operation and scale the ordinate. Finally, the n-order MFCC is calculated by DCT.

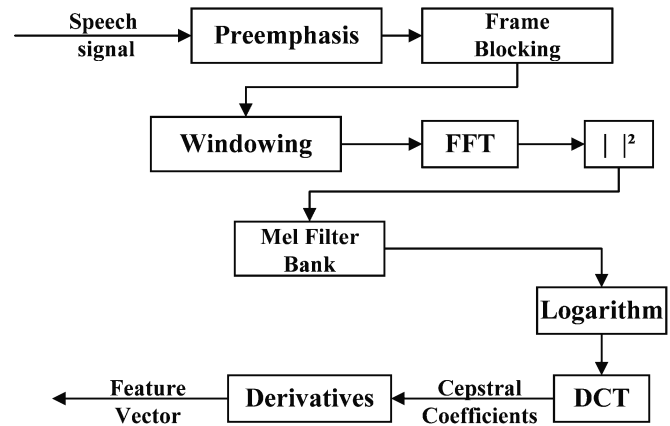

Fig. 2. The overall process of MFCCs

- Pre-emphasis (increase the high-frequency resolution of speech)
  $$signal[i] = signal[i + 1] - a * signal[i] \quad (1)$$
  $a$: pre-emphasis coefficient, taken as 0.97.
- Framing (collect sampling points into one observation unit)

$$frame\ num\ =\ \frac{(N-win+inc)}{inc} \quad (2)$$

- Windowing (solve trailing phenomenon of spectrum in frequency band)

$$x_1[i] = x_0[i] * w[i] \quad (3)$$

Multiply the sampling point of each frame with corresponding element of the hamming window to make the global more continuous and avoid the Gibbs phenomenon of discontinuous boundary.

- Fast Fourier Transformation (*FFT*)

$$X[k] = \sum_{n=0}^{N-1} x[n]e^{-j\left(\frac{2\pi}{N}\right)kn} \quad (4)$$

After FFT, an array with an output dimension of N/2+1 is obtained. Each element of the array represents the energy of the voice signal in a frame and can also express the characteristics of different voices. At the same time, the energy spectrum of the signal is obtained by taking the square of the mode of the spectrum.

- Mel filter bank

The spectrum is smoothed to eliminate the effect of harmonics and highlight rh resonance peak of the original voice. It is worth noting that this operation shows that the speech recognition model based on MFCC will not be affected by the pitch of input voice.

$$Y(m) = \sum_{k=1}^{N} H_m(k)|X[k]|^2 \quad (5)$$

$H_m(k)$: frequency response

- Mel spectrogram

By passing the energy spectrum through the Mel filter, the Mel spectrogram as shown in Fig. 3 can be obtained. Mel spectrogram figures describe the amplitude of the audio stream at various frequencies in a more vivid way. The amplitude of each frequency is shown in the form of a heatmap, the brighter the color is, the higher the energy of the signal is, which indicates the concentration of the sound sample in a specific frequency range. The four pictures in figure 3 show the /a/ sound Mel spectrogram of selected patients with four voice diseases. Theoretically, we can tell the difference between each sample from such a figure, if the resolution is high enough. But in fact, this can only be regarded as a visual description, which still needs objective judgment from a mathematical point of view.

- Discrete Cosine Transform (decorrelation and to strip the low-frequency part)

$$s(m) = log\left(\sum_{k=0}^{N-1}|X(k)|^2 H_m(k)\right) \quad (6)$$

$0 \leq m \leq M$ $\quad M$: the number of filters

$$c(n) = \sum_{m=0}^{N-1} s(m)\cos[\frac{\pi n}{M}\left(m-\frac{1}{2}\right)] \quad (7)$$

n=1,2,3,4…L $\quad$ L: the order of MFCC

As the standard practice, the first 13 MFCCs will be kept, the rest will be discarded [32]. We try to choose 13, 40, 128 MFCCs for experiments (including static MFCC and first order, second order, and higher-order differential coefficients) to represent the static and dynamic characteristics of speech signals effectively and comprehensively. Finally, we extracted a two-dimensional input feature.

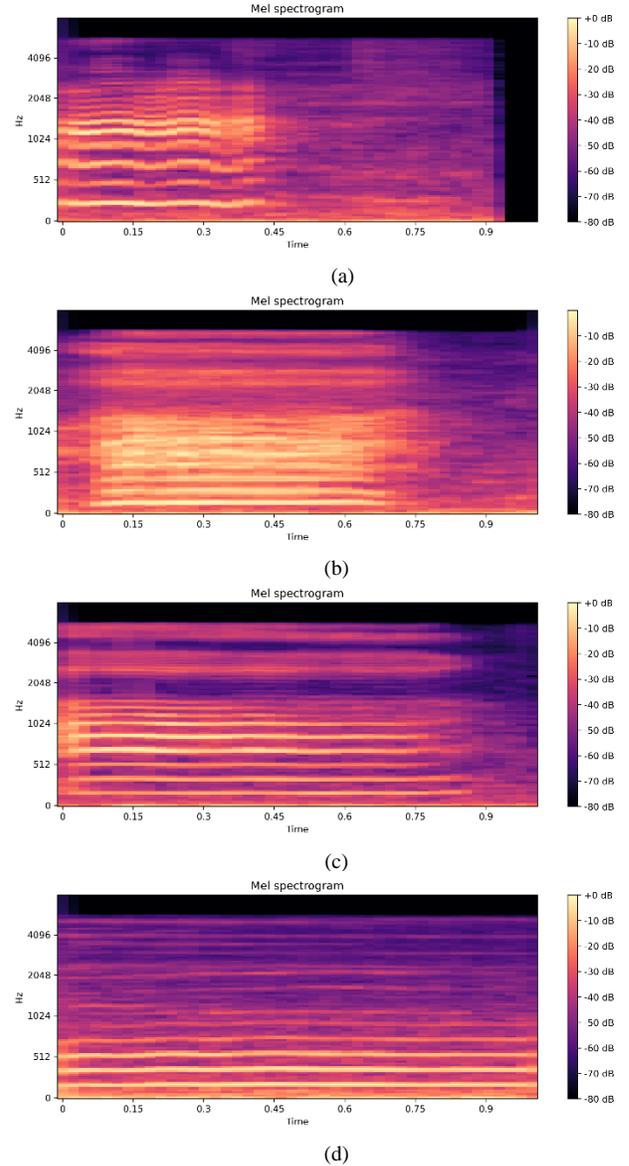

Fig. 3. Mel spectrogram of four pathological samples. (a)Mel spectrogram of patient with Spasmodic dysphonia; (b)Mel spectrogram of patient with Vocal cord paralysis; (c) Mel spectrogram of patient with Vocal cord nodules; (d) Mel spectrogram of patient with Vocal cord polyps.

### C. CNN Model

The research goal of this paper is to classify the voice of patients with voice diseases and identify what kind of disease it is. The model input is the processed MFCCs feature tensor, and the output is the disease label vector. We hope that the performance of the model includes simplicity, efficiency, and high accuracy. Therefore, we have studied a learning-based model, and the structure is shown in Fig. 4.

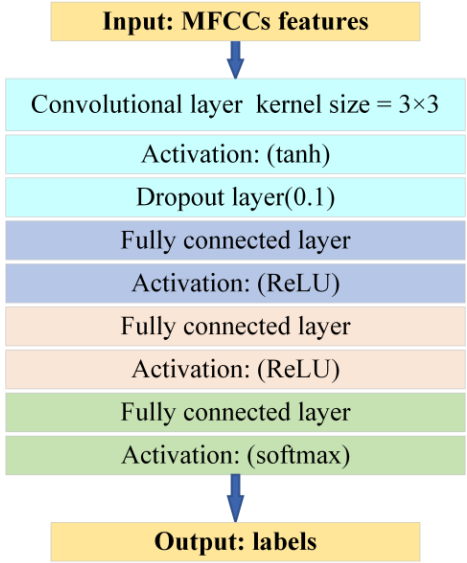

Fig. 4. CNN model architecture

In the problem of image classification, there are many "champion" models, all of which without exception have very complex deep network structures, such as GoogleNet, VGG, ResNet, etc. We also try to apply them to the classification of voice diseases in this paper, and the results are not satisfactory, and the accuracy and recall rate are not high.

The core idea of CNN is sparse interaction, and the selective perception of pre-layer data by multi-layer neurons is disadvantageous in speech data processing in this paper. Feature data are constantly compressed, mapped, and multi-layer convolution operations pay more attention to the local correlation of features, and lack of understanding and global expression of features. We tried several structures of multi-layer convolutional network, first using 5-layer 2D convolutional layers, which kernel size is 3×3. At the same time, we tried to change the size of the kernel function (such as 5 × 5 and 7 × 7), and the number of layers. Finally, it was found that one layer of convolution with a 3×3 kernel function achieved satisfactory results. At the same time the Universal Approximation Theorem states that a feed-forward network with a single hidden layer containing a finite number of neurons, can approximate continuous functions on compact subsets of n [33]. We believe that the network with one hidden layer can approximate the classification function we want. Formula (8) shows that input $X$ and convolution kernel function $k$ do cross-correlation operations, and a scalar bias is added to obtain the output $Y_0$ of the convolution layer.

$$Y_0 = X * k + b_0 \qquad (8)$$

It should be noted that activation function 'tanh' is generally used in RNN network structures, and 'ReLU' function is commonly used in CNN. In multi-layer convolution neural network, the outputs of 'ReLU' function have both >1 and <1 values. After multiple layers of superposition and multiplication, several values greater than 1 and less than 1 compensate for each other. In a sense, this counteracts the possibility of gradient explosion. However, if there is only one layer of convolution, the above compensation situation will not appear, using 'ReLU' function will result in non-convergence as Fig. 5 shows. At the same time, 'ReLU' function will cause the death of neurons due to its mathematical characteristics. We also experimented with 'Leaky ReLU' function. The results were improved, but there were still fluctuations. Using activation function 'tanh' can effectively solve the problem of fluctuations.

$$Y_1 = tanh(Y_0) = \frac{e^{Y_0} - e^{-Y_0}}{e^{Y_0} + e^{-Y_0}} \qquad (9)$$

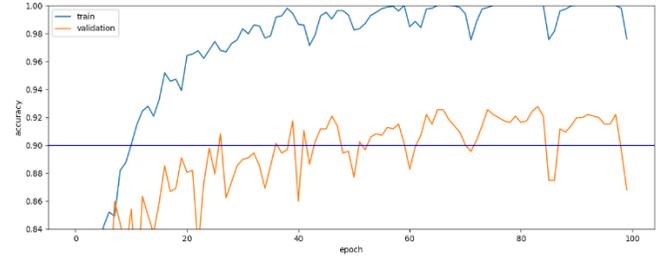

Fig. 5. Accuracy variation of ReLU activation

For the samples we trained, the training model is easy to over-fit due to the dependence on local features. The dropout layer temporarily discards half of the hidden neurons with probability p, updates the parameters of the undeleted neurons, and repeatedly operates as Fig. 6 shows. The picture on the left shows the general layer structure, and the circle with × on the right represents the temporarily discarded neurons. This operation is equivalent to averaging the over-fitting (including forward fitting and reverse fitting) of different neural networks [34]. It also reduces the interaction of hidden nodes with fixed relationships, like the effect of L1 and L2 regularization, and enhances the robustness of the network. The operation of dropout layer also increases the sparsity of features and improves the generalization ability.

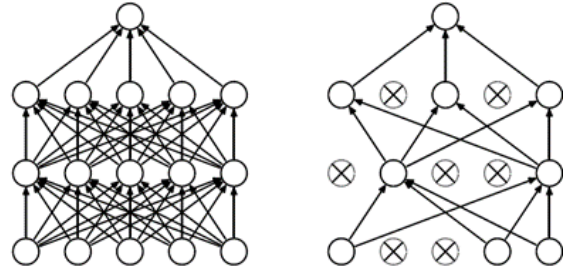

Fig. 6. Schematic diagram of dropout layer

$$r_j \sim Bernoulli(p = 0.1)$$
$$Z_1 = W_1(r_j * Y_1) + b_1 \qquad (10)$$

Flatten the multidimensional object to one dimension (1×8192), then we use three full connection layers to map the features and reduce the final outputs to 4 (the number of final classification labels). The last linear layer is activated by 'softmax' function, the others are activated by 'ReLU'. The mapping process is shown in Fig. 7, and Formula 11 to 13 show the operation process.

$$H_1 = ReLU(Z_1 W_2 + b_2) \qquad (11)$$
$$H_2 = ReLU(H_1 W_3 + b_3) \qquad (12)$$
$$O = softmax(H_2 W_4 + b_4) \qquad (13)$$

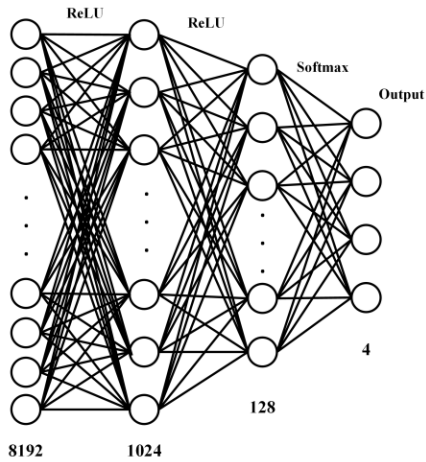

Fig. 7. Architecture of full connection layers

## V. RESULTS

During machine learning training, we should not only train the model, but also use the data to adjust its super parameters and evaluate the model capability. Here, we randomly divide the dataset into three parts: training set, validation set and test set. 80% of the dataset is used to train the model in the training set, 10% is used to verify the set to determine the optimal network structure, and the last 10% is used to evaluate the effect of the model.

When evaluating the effect of the model, we selected three common evaluation indicators: accuracy, recall, and f1-score [35]. At the same time, combing ROC images, the model is evaluated objectively from the perspective of 'accurate' and 'complete.'

First, we started experimenting with the number of MFCCs, find the best number among 13, 40, 50, 128. The results show in Table II.

Table II.
Accuracy of different number of MFCCs

| MFCCs | 13 | 40 | 50 | 128 |
|---|---|---|---|---|
| Accuracy | 73% | 83% | 85% | 92% |

After finding the most suitable output feature parameters, we describe the overall accuracy and loss of the model, as shown in Fig. 8 and Fig. 9.

In Fig. 8, we can see that with the growth of epoch, the training accuracy of the model tends to be stable at 1, and the test accuracy gradually stabilized at about 0.92 after rising. In Fig. 9, the model cross entropy loss of the training set is close to 0.01, and the test set loss is also stable at about 0.3. And the curve of loss conforms to the normal trend of change, there is no over-fitting, insufficient learning, etc.

The confusion matrix is usually used as a first-level indicator. As shown in Fig. 10, the classification performance of the model is directly described with the number of samples classified by the model and the number of samples of the actual category as horizontal and vertical coordinates. Through the representation of the heat map, it is very intuitive to see the difference between the prediction and the actual value of each category.

The confusion matrix counts the number of samples, it counts the number of real and predicted samples. In Fig. 10 we can see values of the correct and incorrect predictions for each class, there are several misjudged audios in the confusion matrix. However, in practical applications, we will vote multiple audios of one patient. The one with the largest number of votes among the four categories is the final predicted disease category. The actual diagnosis accuracy rate should exceed 92%, reaching 98% to almost 100%, which is also the difference between practical application and academic research.

We then calculate the underlying indicators to get the secondary indicator results as shown in Table III.

Table III
Performance on the voice dataset

| Classes | precision | recall | f1_score | support |
|---|---|---|---|---|
| Spasmodic dysphonia | 0.9252 | 0.9519 | 0.9384 | 104 |
| Vocal cord paralysis | 0.8230 | 0.8158 | 0.8194 | 114 |
| Vocal cord nodules | 0.7624 | 0.8280 | 0.7938 | 93 |
| Vocal cord polyps | 0.9018 | 0.8632 | 0.8821 | 234 |

For the four labels that we need to multi-classify, we set a threshold value of 0.5 when classifying the model. So, we worked out the ROC curve, as shown in Fig. 11, to reflect the sensitivity of the model to each label. At the same time, we calculate the figure area under the ROC curve (ROC-AUC-score), which reflects the effect of the classifier. The higher the value of AUC, the better the evaluation of the classifier.

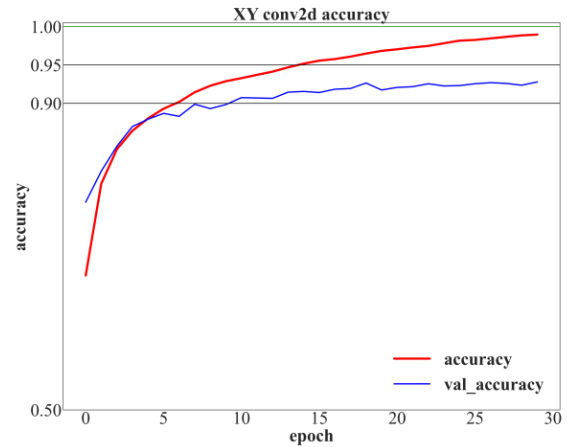

Fig. 8. Accuracy of model on XY dataset

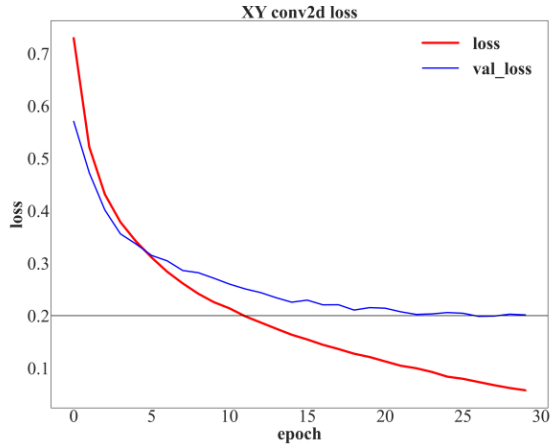

Fig. 9.  Loss of model on XY dataset

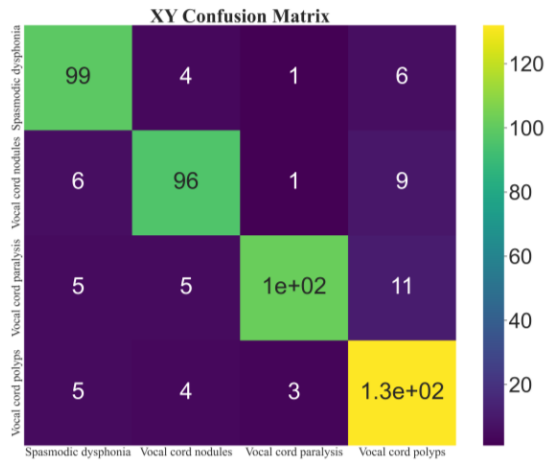

Fig. 10.  Confusion matrix of model on XY dataset

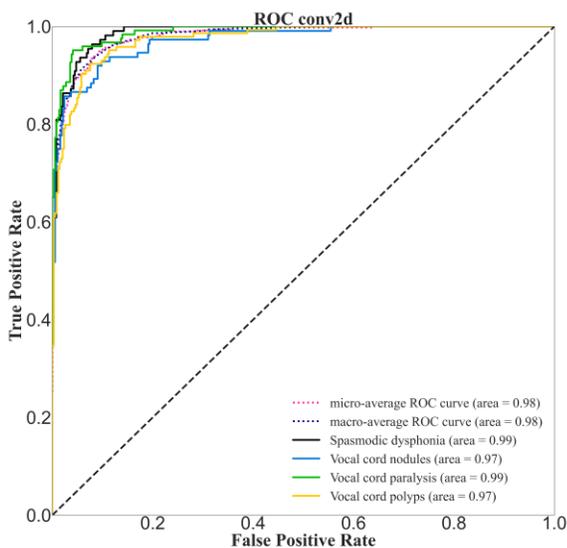

Fig. 11.  ROC curve of model on XY dataset

## VI. Discussion

The use of data-based automated models can provide clinicians with objective judgment assistance. We can see from the model test results in the previous section that the accuracy rate reached 92%. It is also worth mentioning that our model performs fast when dealing with voice and audio. Most of the detection and classification systems related to voice diseases use accuracy, specificity, and sensitivity as evaluation indicators. However, the evaluation based on execution time is also important, especially when the system of voice pathology detection and classification is applied clinically. The long waiting time is likely to affect the patient's emotional and psychological state, and patients' trust in computer-aided diagnostic tools will also be reduced.

In addition, we believe that our method also has strong universality, that is, the generalization ability is dependable. We use The Advanced Voice Function Assessment Databases (AVFAD) as an object to evaluate the generalization ability of the model. We conducted several related experiments in various conditions. The experimental results show in Table IV-VI.

1) AVFAD extraction of 13, 49, 50, 128 MFCCs
2) Monophonic or polyphonic combination as input, we tried /i/, /a, i/, /a, i, u/, three conditions, and repeated for 3 times.

Table IV.
Accuracy under different MFCCs using ternary vowels

| MFCCs    | 13  | 40  | 50  | 128 |
|----------|-----|-----|-----|-----|
| Accuracy | 92% | 98% | 98% | 98% |

Table V
Accuracy under different MFCCs using binary vowels

| MFCCs    | 13  | 40  | 50    | 128 |
|----------|-----|-----|-------|-----|
| Accuracy | 91% | 97% | 97.5% | 98% |

Table VI
Accuracy under different MFCCs using single vowel

| MFCCs    | 13  | 40  | 50  | 128 |
|----------|-----|-----|-----|-----|
| Accuracy | 90% | 97% | 97% | 98% |

The input audio using ternary vowels as analysis has a better classification effect, with an accuracy of about 98%. The following are the images of the accuracy curve changing with the increase of MFCCs.

The higher the MFCCs value is, the higher the detected accuracy is. For the big data set (AVFVD), it reaches a peak at about 50 and then tends to remain unchanged. However, the situation of curve oscillation is still obvious, and the MFCCs with larger value (128) is improved, that is, the stability is better, and the fitting efficiency is also improved as Fig.12-15 show.

At the same time, when it comes to the Xiangya dataset, we find that the small dataset needs a larger MFCCs to achieve better results, while the big data set takes a moderate value to achieve high accuracy.

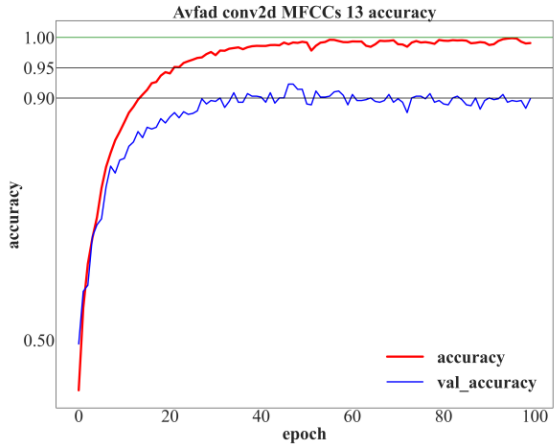

Fig. 12.  Accuracy of model on AVFAD with 13MFCCs

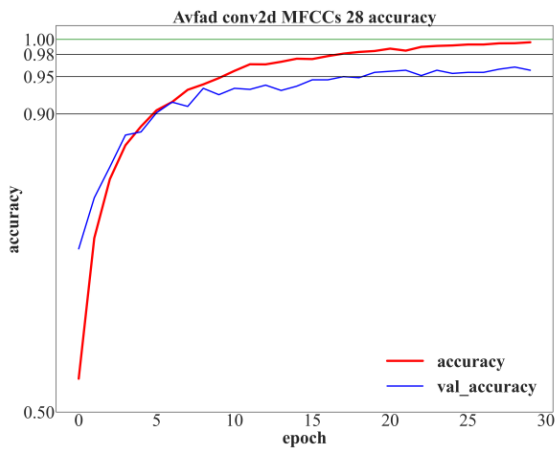

Fig. 13.  Accuracy of model on AVFAD with 28MFCCs

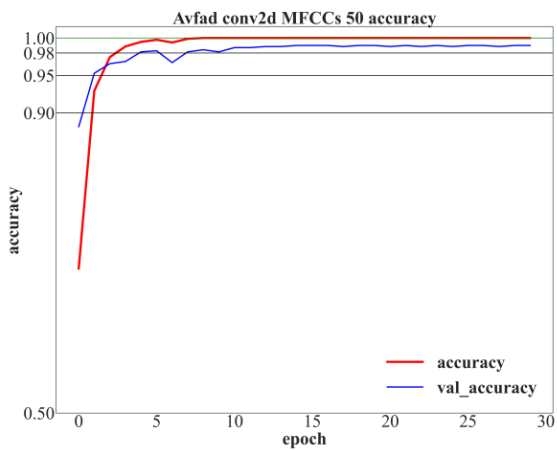

Fig. 14.  Accuracy of model on AVFAD with 50MFCCs

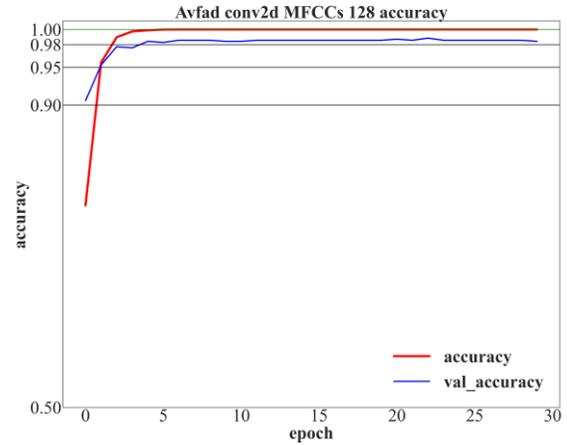

Fig. 15.  Accuracy of model on AVFAD with 128MFCCs

## VII. CONCLUSION

In order to meet the clinical needs of Xiangya Hospital and improve the diagnostic efficiency of voice diseases, we propose a more accurate and rapid method to classify four common voice diseases. First, we used the normal quality patient audio (8 phonemes) collected in the clinical environment to train the model, adjusted the parameters, and tested the classification effect with an accuracy of 92% and a recall rate of 86%. To prove the universality of the method, we verified it on the AVFAD, and the accuracy is more than 98%.

The contributions of this paper are mainly reflected in the following aspects:

1) Most of the pathological studies on voice diseases focus on the two categories of healthy speech and pathological speech, and the audio of the study is also in the unit vowel /a/. In this paper, 8 phonemes are collected to classify the four categories of the disease itself.
2) Different from most CNN-based speech classification models, we propose a single convolution layer structure, which proves that it has more accurate and faster classification effect.
3) In the experiment, we get the general rule of speech feature MFCC. Big data set takes a moderate MFCCs value, and a small data set takes a larger MFCCs value, which can make up for the defects of input features and achieve good classification results.

We believe that the patient information in Xiangya data set is also valuable for research, such as working environment, eating habits and so on, which can be used as the focus of auxiliary voice disease research. Adding these factors to the study can make the prediction more forward-looking. In addition, using the two-dimensional description of the Mel diagram in figure 3, we believe that the image of each patient is also specific, from which we can also learn features, and applying advanced image recognition methods here is also our direction.

## ACKNOWLEDGMENT

We sincerely thank the Otorhinolaryngology team of Xiangya Hospital for collecting such valuable speech data and providing professional medical knowledge. Additionally, we would thank University of Aveiro for providing us AVFAD to verify our model.